\documentclass[12pt,letterpaper]{JHEP3}
\usepackage{epsfig}
\def\Box{\square}

\def \tilde{\widetilde}

\def \eqref#1{(\ref{#1})}

\def \ignoruj#1{}
\def \eqn#1#2{\begin{equation}#2\label{#1}\end{equation}}

\def \half{{1\over 2}}

\def\ignorethis#1{}
\def\be{\begin{equation}}
\def\ee{\end{equation}}

\def\bear{\begin{eqnarray}}
\def\eear{\end{eqnarray}}

\def\t#1{\mathrm{#1}}

\def\inf{\infty}
\def\->{\rightarrow}
\def\=={\begin{eqnarray*}}
\def\xx{\end{eqnarray*}}
\def\n==#1{\begin{eqnarray} \label{#1}}
\def\nxx{\end{eqnarray}}
\def\half{\frac{1}{2}}
\def\shalf{\textstyle\frac{1}{2}\displaystyle}
\def\sfrac#1#2{\textstyle\frac{#1}{#2}\displaystyle}

\title{Higher-order corrections to mass-charge relation of extremal black holes}

\author{Yevgeny Kats, Lubo\v{s} Motl and Megha Padi\\
  Jefferson Physical Laboratory\\
  Harvard University, Cambridge, MA 02138, USA\\
E-mail: \email{kats, motl, padi at fas.harvard.edu}}

\abstract{We investigate the hypothesis that the higher-derivative
corrections always make extremal non-supersymmetric black holes
lighter than the classical bound and self-repulsive. This
hypothesis was recently formulated in the context of the so-called
swampland program. One of our examples involves an extremal
heterotic black hole in four dimensions. We also calculate the effect of general four-derivative terms in Maxwell-Einstein theories in $D$
dimensions. The results are consistent with the conjecture.
}


\preprint{{\tt hep-th/0606100}\\HUTP-06/A0023}
\begin{document}


\section{Introduction\label{sec-Intro}}

In view of the seemingly large number of allowed vacua in string theory,
it is important to look for universal properties of these solutions, and
see what features of the low-energy field theory can nevertheless be
deduced from string theory. It turns out that such features exist, and not
any low-energy particle content is allowed: Vafa \cite{swampland} has
discussed the possibility of restrictions related to the finiteness of
volume of massless scalar fields, the finiteness of the number of massless fields, and the rank of the gauge groups. In fact, just the requirement to include quantum gravity (even if not in the framework of string theory) puts constraints on the low-energy physics
\cite{weak-grav,o-vafa-swamp,Li-Song-Wang,Huang-Li-Song,Li-Song-Song-Wang, Banks,Huang-grav-corr,Huang-infl,Eguchi-Tachikawa,Huang-spect-ind}.
Arkani-Hamed et al.\ \cite{weak-grav} considered a theory of a single U(1)
gauge field, and came to the conclusion that the gauge force must be
stronger than gravity, i.e., there must exist charged particles for which the net force is repulsive. Furthermore, the effective theory breaks down at some scale {\it beneath} the Planck scale, and there should exist a charged particle at or below that scale.

In particular, Arkani-Hamed et al.\ made a prediction regarding the
mass-charge relation of extremal black holes. Consider a particle with a
mass $M$ and a charge $Q$. For this particle to be unstable, it must be
able to decay into two or more particles whose total mass is smaller than $M$ and total charge equal to $Q$. To satisfy these conditions, at least one of the outgoing
particles must have a smaller $M/Q$ ratio than the original particle.

The argument extends to black holes, which are believed to be the
low-energy description of elementary particles whose masses are much above
the Planck scale. Since it is unnatural to have an infinite number of
exactly stable particles, the mass-charge relation for extremal black
holes $M=Q$
cannot be exact: the $M/Q$ ratio for extremal black holes should decrease
with decreasing $Q$, so that for every extremal black hole there is
another black hole with a smaller $M/Q$ ratio (see figure \ref{M-Q-fig}).
Because states with $M/Q < 1$ must exist, the most natural expectation is that the black holes, states with very high values of $M,Q$, also satisfy $M/Q < 1$, although the difference from $1$ is tiny.

\EPSFIGURE{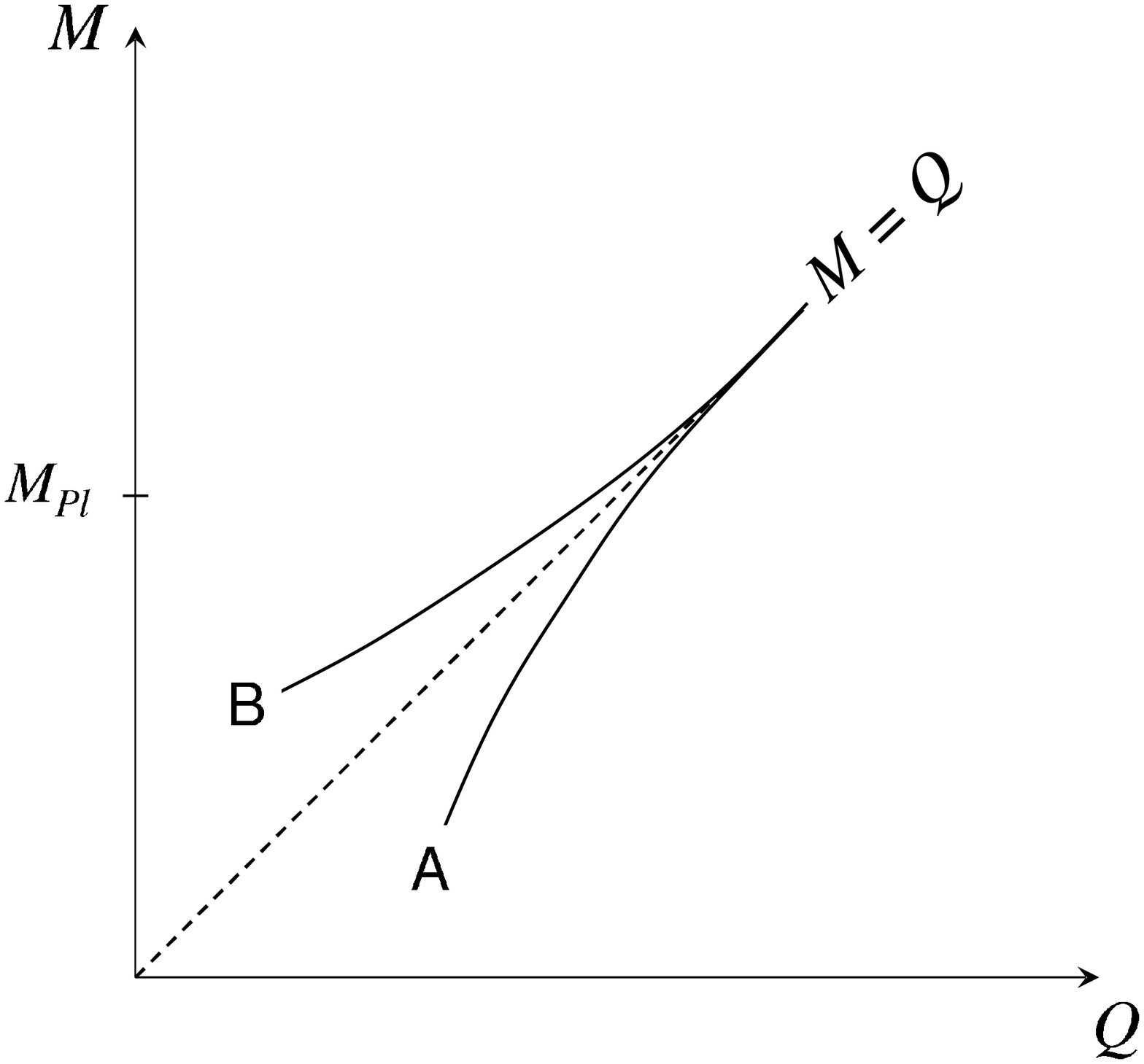,width=120mm}{The classical mass-charge relation for
extremal black holes is represented by the dashed line; it must be valid in the limit $M \gg M_\mathrm{Pl}$. Curve A shows a possible exact mass-charge relation. Curve B is unacceptable because it would imply an infinite number of states that cannot decay.
\label{M-Q-fig}}

Since the net force between black holes with $M = Q$ vanishes, the
previous argument also predicts that the net force will become
repulsive. This is indeed expected because if the force were
attractive, heavier bound states with a lower $M/Q$ ratio would be
possible, again creating an infinite number of states that cannot
decay. While the relation between the decrease of the mass and the
repulsion is trivial in the case of Reissner-Nordstr\"{o}m black
holes, the existence of other fields (e.g., the dilaton) makes the
two arguments independent.

In this paper we present calculations concerning corrections to
the mass-charge relation of extremal black holes. Section
\ref{sec-RN} is dedicated to the case of four-derivative terms
affecting Reissner-Nordstr\"{o}m black holes (and Appendix
\ref{app-ddim} extends the result to the case of $D$ dimensions).
Section \ref{sec-GHS} discusses a heterotic black hole where the
additional coupling to the dilaton must be included. In Section
\ref{sec-discussion}, we offer conclusions and a list of black
objects that could be investigated.

\section{Corrections to the Reissner-Nordstr\"{o}m black hole\label{sec-RN}}

The Reissner-Nordstr\"{o}m black hole is a spherically symmetric
static solution with a radial electric (or magnetic) field, governed
by the action: \eqn{RN-action}{ S = \int d^4x
\sqrt{-g}\left(\frac{R}{2\kappa^2} -
\frac{1}{4}F_{\mu\nu}F^{\mu\nu}\right) } where $\kappa^2 = 8\pi G$.
Starting with the most general spherically symmetric static metric
\eqn{gen-metric} {ds^2 = -e^{\nu(r)} dt^2 + e^{\lambda(r)} dr^2 +
r^2 d\Omega^2} and looking for a solution with a radial electric
field of the form $F^{01} = E(r)$, one finds \eqn{RN-metric}
{e^{\nu(r)} = e^{-\lambda(r)} = 1 - \frac{\kappa^2 M}{4\pi r} +
\frac{\kappa^2 Q^2}{32\pi^2 r^2} \qquad\qquad E(r) = \frac{Q}{4\pi
r^2}} The solution describes a black hole for $M \geq
\frac{\sqrt{2}}{\kappa}\,|Q|$ (otherwise the solution describes a
naked singularity). Black holes with the minimal possible mass $M$
for a given charge $Q$ are called extremal. In units with $\kappa^2
= 2$, they satisfy $M = |Q|$ and the horizon radius $r = M/4\pi =
|Q|/4\pi$.

Corrections due to quantum gravity can be represented by higher-order terms in the effective action. For the purpose of determining the mass of an extremal black hole, we are interested in the solution near the horizon: $r \sim Q$. The unperturbed solution \eqref{RN-metric} implies that any derivative contributes a factor of order $1/Q$, so the Riemann tensor is $R \sim 1/Q^2$, and for the electromagnetic field tensor we have $F \sim Q/r^2 \sim 1/Q$ and $\nabla F \sim 1/Q^2$. Since $Q \simeq M \gg 1$, terms of higher order in $R$, $F$, and derivatives are suppressed by powers of $1/Q$, and we may consider just the leading-order corrections.
Both terms in \eqref{RN-action} are $\sim 1/Q^2$. The leading order
($\sim 1/Q^4$) corrections are: \n=={RN-action-corr} S = \int d^4x
&& \sqrt{-g} \left( \frac{R}{2\kappa^2} -
\frac{1}{4}F_{\mu\nu}F^{\mu\nu}
+ c_1\,R^2 + c_2\, R_{\mu\nu}R^{\mu\nu} + c_3\,R_{\mu\nu\rho\sigma}R^{\mu\nu\rho\sigma} + \right. \nonumber \\
&& \left. + \, c_4\,RF_{\mu\nu}F^{\mu\nu} +
c_5\,R^{\mu\nu}F_{\mu\rho}{F_\nu}^\rho +
c_6\,R^{\mu\nu\rho\sigma}F_{\mu\nu}F_{\rho\sigma} +
c_7\,(F_{\mu\nu}F^{\mu\nu})^2 \right. \nonumber \\
&& \left. + \, c_8\,(\nabla_{\mu} F_{\rho\sigma})(\nabla^{\mu}
F^{\rho\sigma}) + c_9\,(\nabla_{\mu} F_{\rho\sigma})(\nabla^{\rho}
F^{\mu\sigma}) \right) \nxx We did not include a $(\nabla_\mu
F^{\mu\nu})(\nabla^\rho F_{\rho\nu})$ term because $(\nabla_\mu
F^{\mu\nu})$ and $(\nabla^\rho F_{\rho\nu})$ vanish in the
unperturbed solution, so variations of this term are proportional to
additional powers of the correction coefficients $c_i$. A similar
argument applies to $\tilde{c}_7
F^{\mu\nu}F_{\nu\rho}F^{\rho\sigma}F_{\sigma\mu}$, whose
contribution to the equations of motion (to first order in $c_i$)
turns out to be equal to half the contribution of
$(F^{\mu\nu}F_{\mu\nu})^2$, related to the fact that only $F^{01}$
and $F^{10}$ are non-zero in the unperturbed solution. Therefore, in our problem
$\tilde{c}_7$ can be absorbed in $c_7$.

The solution of the equations of motion for the metric is
straightforward \cite{RN-corr}. First, one can note that the
spherical symmetry made it possible to express $\lambda(r)$ and
$\nu(r)$ explicitly in terms of $R_{\mu\nu}$ as
\eqn{lambda-R}{e^{-\lambda} = 1 - \frac{\kappa^2 M}{4\pi r} -
\frac{1}{r}\int_r^\inf dr\,r^2 \left( \frac{R_0^0 - R_1^1}{2} -
R_2^2 \right) } \eqn{nu-R}{ \nu = - \lambda + \int_r^\inf dr\,r
\left(R_0^0 - R_1^1\right)\,e^\lambda } Next, recall Einstein's
equation in the form \eqn{Einstein}{ R_{\mu\nu} = \kappa^2
\left(T_{\mu\nu} - \shalf Tg_{\mu\nu} \right) \qquad\qquad
T_{\mu\nu} = -\frac{2}{\sqrt{-g}} \frac{\delta
S_\t{matter}}{\delta g^{\mu\nu}} } where $T = T^0_0 + T^1_1 +
T^2_2 + T^3_3$ (with $T^3_3 = T^2_2$). Then \eqref{lambda-R} and
\eqref{nu-R} become \eqn{lambda}{ e^{-\lambda} = 1 -
\frac{\kappa^2 M}{4\pi r} - \frac{\kappa^2}{r}\int_r^\inf
dr\,r^2\,T_0^0 } \eqn{nu}{ \nu = - \lambda + \kappa^2 \int_r^\inf
dr\,r \left(T_0^0 - T_1^1\right)\,e^\lambda } We take the
higher-order terms in the action \eqref{RN-action-corr} to be a
perturbation, treat them as a part of $S_\t{matter}$, and use the
unperturbed solution \eqref{RN-metric} to calculate their
corresponding $T_{\mu\nu}$. We also vary the action with respect
to the gauge field to obtain corrections to Maxwell's equations,
which modify the contribution of the
$-\sfrac{1}{4}F_{\mu\nu}F^{\mu\nu}$ term to $T_{\mu\nu}$. The
calculation of these two contributions to the effective
$T_{\mu\nu}$ is presented in Appendix \ref{app-Tmn}. The corrected
metric in terms of $m=M/4\pi$ and $q=Q/4\pi$ is
\n=={CorrectedMetric} e^{-\lambda}&=&1-\frac{\kappa^2
m}{r}+\frac{\kappa^2 q^2}{2r^2}+ \frac{q^2}{r^6} \left(c_2
\frac{\kappa^4}{5}\left(-6\kappa^2 q^2+15m \kappa^2
r-20r^2\right)\right. \nonumber\\ &&\left. +\, c_3
\frac{\kappa^4}{5} \left(-24\kappa^2 q^2+60\kappa^2
mr-80r^2\right)+ c_4 \kappa^2 \left(-6\kappa^2 q^2+14\kappa^2
mr-16r^2\right) \right.
\nonumber\\&&\left.+\,c_5\frac{\kappa^2}{5} \left(-11\kappa^2
q^2+25\kappa^2 mr-30r^2\right) +c_6\frac{\kappa^2}{5}
\left(-16\kappa^2 q^2+35\kappa^2 mr-40r^2\right) \right.
\nonumber\\&&\left.+\,c_7\left(\frac{-4\kappa^2 q^2}{5}\right) +
c_8 \frac{\kappa^2}{5} \left(6\kappa^2 q^2-15 \kappa^2 mr+20r^2
\right) \right. \nonumber\\&&\left.+\, c_9 \frac{\kappa^2}{10}
\left(6\kappa^2 q^2-15 \kappa^2 mr+20r^2 \right) \right) \nxx The
mass-charge relation for extremal black holes becomes
\n=={RN-mass-corr}
\frac{\kappa}{\sqrt{2}}\frac{M}{|Q|}=1-\frac{2}{5q^2}\left(2c_2 +
8c_3 + \frac{2c_5}{\kappa^2} + \frac{2c_6}{\kappa^2} +
\frac{8c_7}{\kappa^4} - \frac{2c_8}{\kappa^2} -
\frac{c_9}{\kappa^2} \right) \nxx Then the conjecture of
Arkani-Hamed et al.\ implies that our low-energy effective theory
must satisfy \eqn{RN-constraint}{ 2 c_2 \kappa^4 + 8 c_3 \kappa^4
+ 2 c_5 \kappa^2 + 2 c_6 \kappa^2 + 8 c_7 - 2 c_8 \kappa^2 - c_9
\kappa^2 \geq 0 }

We performed the same calculation in $D$ spacetime dimensions, and the
results are presented in Appendix \ref{app-ddim}.

We can use our results to check whether higher-order terms in the
string theory effective action increase or decrease the
mass-charge ratio in certain special cases. A U(1) gauge field can
arise as a subgroup of the E$_8$$\times$E$_8$ or SO(32) gauge
group in the low-energy effective theory of the heterotic string.
We would like to consider a black hole charged under this U(1),
while we set the remaining gauge fields and the antisymmetric
field strength $H_{\mu\nu\rho}$ to zero. Consider heterotic string
theory compactified on a ($10-D$)-dimensional torus. If we are
able to stabilize the dilaton, then one possible background is a
$D$-dimensional Reissner-Nordstr\"{o}m black hole. (A black hole
that involves the dilaton as well is discussed in the next
section.) The ten-dimensional Lagrangian is \cite{Gross-Sloan}:
\n=={HetEffAction} {\cal L} &=& \frac{1}{2\kappa_{10}^2}R -
\frac{1}{4}F_{\mu\nu}F^{\mu\nu} + \frac{\alpha'h}{16\kappa_{10}^2}
\left(R_{\mu\nu\rho\sigma}R^{\mu\nu\rho\sigma}-
4R_{\mu\nu}R^{\mu\nu}+R^2\right) \nonumber \\
&&-\frac{3}{64} \alpha'h \kappa_{10}^2
\left((F_{\mu\nu}F^{\mu\nu})^2-4F^{\mu\nu}F_{\nu\rho}F^{\rho\sigma}F_{\sigma\mu}\right)
\nxx The dilaton has been set to a constant $\phi_0$ and $h \equiv
e^{-\kappa_{10} \phi_0/\sqrt{2}}$. Such an assumption may be physically
interpreted as a consequence of a dynamically generated potential
for the dilaton in a particular compactification: the dilaton
acquires mass much greater than the inverse radius of the black
hole, its effects may be neglected, while the terms we consider
are preserved. In $D=4$, the Gauss-Bonnet combination
$$R_{\mu\nu\rho\sigma}R^{\mu\nu\rho\sigma}-4R_{\mu\nu}R^{\mu\nu}+R^2$$
is a topological invariant and does not influence the equations of
motion. It does have an effect in other dimensions, where it
interestingly cancels the $(3D-7)$ factor in \eqref{mass-ddim}.
While the effect of the Gauss-Bonnet terms is to increase the
mass, the combination of the $F^4$ terms decreases the mass. (Note
also that when the $F^4$ terms are expressed in terms of $(F^2)^2$
and $(F\tilde F)^2$, their coefficients are positive, much like in
the Dirac-Born-Infeld action: this fact is required by the energy
conditions or, equivalently, the unitarity \cite{nima-allan}.)
With \n=={HetCoeffs} c_1 = c_3 = \frac{h\alpha'}{16\kappa^2}\qquad
c_2 = -\frac{h\alpha'}{4\kappa^2}\qquad c_7 = \frac{3 h\alpha'
\kappa^2 }{64} \nxx where we absorbed $\tilde{c}_7$ in $c_7$ as
explained after eq.\ \eqref{RN-action-corr}, we obtain
\n=={HetMassCorrection} \frac{D-3}{D-2} \frac{\kappa^2 M^2}{Q^2}
&=& 1 - \alpha'\frac{(D-3)(2D-5)h}{4(3D-7)}
\left(\frac{(D-2)(D-3)\Omega_{D-2}^2}{\kappa^2
Q^2}\right)^{1/(D-3)} \nxx The overall effect is to lower $M/Q$
for $D > 3$, as we indeed expect for a theory that includes
quantum gravity.

Interestingly, the leading term in $D$ canceled in \eqref{HetMassCorrection}, which might be relevant in large-$D$ expansions. The reader may also notice that the leading mass correction parametrically agrees with the relation for perturbative string excitations only in $D=4$, where both relations can be written as
\n=={ConstShift}
M^2 = a Q^2 - b
\nxx
where $a$ and $b$ are constants.

\section{Corrections to the GHS black hole\label{sec-GHS}}

In general, the low-energy effective action of the heterotic
string includes also the dilaton field $\phi$, which is sourced by
the gauge field: \eqn{GHS-action}{ S = \int d^4x \sqrt{-g}\left(R
- 2(\nabla\phi)^2 - e^{-2\phi}F_{\mu\nu}F^{\mu\nu}\right) } When
the dilaton is present, the Reissner-Nordstr\"{o}m metric is no
longer a solution to the equations of motion. Black holes charged
under a U(1) gauge field must also carry dilatonic charge, as was
analyzed by Garfinkle et al.\ (GHS) \cite{GHS}. A magnetically
charged black hole ($F = Q \sin\theta d\theta \wedge d\varphi$) is
then described by \n=={GHS-soln}
ds^2 &=& - \left(1 - \frac{2M}{r}\right) dt^2 + \left(1 - \frac{2M}{r}\right)^{-1} dr^2 + r \left(r - \frac{Q^2\,e^{-2\phi_0}}{M}\right) d\Omega^2 \\
e^{-2\phi} &=& e^{-2\phi_0} \left(1 - \frac{Q^2\,e^{-2\phi_0}}{Mr}\right)
\nxx
where $\phi_0$ is the asymptotic value of $\phi$ at infinity, which we set to zero, for simplicity. The black hole has a horizon at $r=2M$ for $M > |Q|/\sqrt{2}$. The solution for the dilaton implies that the black hole has a dilatonic charge of $D = -Q^2/2M$, which for the extremal case reduces to $D = -M$. The force between two particles with magnetic charge $Q$, dilatonic charge $D$, and mass $M$, is given by
\eqn{force-w-dilaton}{ F = \frac{Q^2 - D^2 -
M^2}{16\pi r^2} }
so the net force between two extremal black holes with equal charges
vanishes. The argument of Arkani-Hamed et al.\ would then predict that higher-order corrections to the mass and the dilatonic charge would make the mass smaller and the net force repulsive as the charge $Q$ becomes smaller.

Corrections to the metric and dilaton field of a magnetically-charged GHS black hole due to the next order terms ($R^2$, $F^4$, $F^2(\nabla\phi)^2$) in the heterotic string effective action have been calculated by Natsuume \cite{Natsuume}. After eliminating many of the terms by field redefinitions, he obtained the corrections to leading order in $\alpha'$ as
\n=={spherfourteen}
{\cal L} = a\left(R_{\mu\nu\rho\sigma}R^{\mu\nu\rho\sigma} - 4R_{\mu\nu}R^{\mu\nu} + R^2\right) + b (F^2)^2 + c F^2(\nabla\phi)^2 + h R^{\mu\nu\rho\sigma}F_{\mu\nu}F_{\rho\sigma}
\nxx
The coefficients of $R_{\mu\nu\rho\sigma}R^{\mu\nu\rho\sigma}$ and $R^{\mu\nu\rho\sigma}F_{\mu\nu}F_{\rho\sigma}$, which are invariant under field redefinitions, were then taken from the heterotic string calculations \cite{Gross-Sloan}: $a = \alpha'/8$ and $h = 0$. The perturbed equations of motion were written down, and a requirement of consistency with exact results that were obtained for this black hole \cite{GPS} determined $c=\alpha'/2$. The value of $b$ does not affect the correction to the mass. The metric (in the extremal limit) becomes
\n=={Natsuume-metric}
ds^2 = &-& \left(1-\frac{\tilde{Q}}{r}\right)^{1+\epsilon} f_2\left(\frac{\tilde{Q}}{r}\right)
dt^2 + \left(1-\frac{\tilde{Q}}{r}\right)^{-1+\epsilon}f_3\left(\frac{\tilde{Q}}{r}\right)dr^2 + \\ \nonumber
&+& r^2 \left(1-\frac{\tilde{Q}}{r}\right)^{1+\epsilon}f_4\left(\frac{\tilde{Q}}{r}\right)d\Omega^2
\nxx
and the dilaton is given by
\eqn{Natsuume-dilaton}
{e^{-2\phi} = \left(1-\frac{\tilde{Q}}{r}\right)^{1+\epsilon}
f_4\left(\frac{\tilde{Q}}{r}\right) }
where $\tilde{Q}=\sqrt{2}\,Q$, $\epsilon = (2b-1) \alpha'/\tilde{Q}^2$, and
\n=={Polynomials} f_2(x)&=&1-\frac{\alpha'}{40\tilde{Q}^2} x
(11x^3+7x^2+16x+38)+g(x)
\\f_3(x) &=& 1-\frac{\alpha'}{40\tilde{Q}^2} x(19x^3+25x^2+26x+42)+g(x)\\
f_4(x)
&=&1-\frac{\alpha'}{40\tilde{Q}^2}x(-9x^3+7x^2+16x+38)+ g(x) \\
g(x)&=&\frac{\alpha'}{60\tilde{Q}^2}b x (15x^3+32x^2+57x+120) .
\nxx
Natsuume found that the mass-charge relation for the extremal black
holes (with the normalization given in our eq.\ \eqref{GHS-action})
is given by
\eqn{GHS-mass-corr}{
M = \frac{|Q|}{\sqrt{2}} \left( 1 - \frac{\alpha'}{40Q^2} \right). }
This agrees with the expectation that the $M/Q$ ratio decreases as the charge $Q$ becomes smaller.

Furthermore, we can use eq.\ \eqref{Natsuume-dilaton} to determine
the correction to the dilatonic charge $D$. We identify $D$ as the
coefficient of the $1/r^2$ term in $d\phi/dr$ and obtain the
corrected dilatonic charge of the extremal black hole as
\eqn{GHS-dil-corr}{ D = -\frac{|Q|}{\sqrt{2}} \left( 1 -
\frac{\alpha'}{40Q^2} \right). } Since both the mass and the
dilatonic charge decrease, the net force \eqref{force-w-dilaton}
between the extremal black holes becomes repulsive, as was
conjectured in Section \ref{sec-Intro}.

\section{Discussion\label{sec-discussion}}

We have calculated the corrections to the masses of extremal black
holes in several backgrounds. In all examples where we could
verify the sign, the sign was negative. This fact was not
guaranteed by the general rules of effective field theory;
however, general arguments exist why such an inequality could
follow from the consistency of couplings in quantum gravity
\cite{weak-grav}.

Other examples of black objects where the inequality could be
checked include non-supersymmetric black holes in type II string
theory on Calabi-Yau manifolds and various black branes. It is
desirable to find either a more general proof that the extremal
black holes become lighter in general backgrounds of quantum
gravity or a counterexample. We also conjecture that the first
correction to the Bekenstein-Hawking entropy, arising from
higher-derivative terms applied to Wald's formula, is positive in
all cases. We are not aware of counterexamples; explicit checks or
a more general proof could shed some light on the UV-IR relations
in quantum gravity.

\acknowledgments

\vspace{-3mm}

{\small We are grateful to Allan Adams, Nima Arkani-Hamed, Monica
Guica, Alberto Nicolis, and Cumrun Vafa for useful discussions.
This work was supported by a DOE OJI award and the DOE under
contract DE-FG02-91ER40654. The work of MP was supported by a
National Science Foundation Graduate Fellowship.}

\appendix

\section{Energy-momentum tensor\label{app-Tmn}}

First order corrections to the energy-momentum tensor $T_{\mu\nu}$
have two contributions: a correction to the energy-momentum tensor
of the $-\sfrac{1}{4}F_{\mu\nu}F^{\mu\nu}$ term due to corrections
to $F^{\mu\nu}$, and an effective contribution representing the
modification of Einstein's equation by all the higher-order terms.

To find the first contribution, we vary the action with respect to
$A_\mu$ to obtain the corrected Maxwell's equations:
\n=={Maxwell-corr} \nabla_\nu F^{\mu\nu} = && 4 c_4 \nabla_\nu
(RF^{\mu\nu}) + 2 c_5 \nabla_\nu (R^{\mu\rho}{F_\rho}^\nu -
R^{\nu\rho}{F_\rho}^\mu) +
\\ \nonumber
&+& 4 c_6 \nabla_\nu (R^{\alpha\beta\mu\nu}F_{\alpha\beta}) + 8
c_7 \nabla_\nu(F_{\rho\sigma}F^{\rho\sigma} F^{\mu\nu}) - 4 c_8
\nabla_\nu \Box F^{\mu\nu} -
\\ \nonumber
&-& 2c_9 \nabla_\nu\nabla_\rho(\nabla^\mu F^{\rho\nu} - \nabla^\nu
F^{\rho\mu}) \nxx We find the first-order correction to
$F^{\mu\nu}$ by treating the right hand side as a perturbation (evaluated with the unperturbed metric and electric field). Since $T_{\mu\nu}$ is quadratic in the fields, only corrections to $F^{01}$ (which is non-zero in the unperturbed solution) are of the first order in $c_i$.

To find the second contribution to $T_{\mu\nu}$, we calculate the
variation of the higher-order terms in the action with respect to
$g^{\mu\nu}$, which gives
\n=={Einstein-corr}
\Delta T_{\mu\nu}
&=& c_1 \left(g_{\mu\nu} R^2 - 4RR_{\mu\nu} +
4\nabla_\nu\nabla_\mu R - 4g_{\mu\nu}\Box R \right) + \\ \nonumber
&+& c_2 \left(g_{\mu\nu} R_{\rho\sigma} R^{\rho\sigma} +
4\nabla_\alpha\nabla_\nu R^\alpha_\mu - 2\Box R_{\mu\nu} -
g_{\mu\nu} \Box R - 4R^\alpha_\mu R_{\alpha\nu} \right) + \\
\nonumber
&+& c_3 \left(g_{\mu\nu} R_{\alpha\beta\gamma\delta}R^{\alpha\beta\gamma\delta} -
4R_{\mu\alpha\beta\gamma}{R_\nu}^{\alpha\beta\gamma} - 8\Box
R_{\mu\nu} + 4\nabla_\nu\nabla_\mu R + 8R_\mu^\alpha R_{\alpha\nu}
- 8R^{\alpha\beta}R_{\mu\alpha\nu\beta} \right) + \\ \nonumber
&+& c_4 \left(g_{\mu\nu}R F^2 - 4R{F_\mu}^\sigma F_{\nu\sigma} -
2F^2R_{\mu\nu} + 2\nabla_\mu\nabla_\nu F^2 - 2g_{\mu\nu}\Box
F^2\right) + \\ \nonumber
&+& c_5 \left(g_{\mu\nu}R^{\kappa\lambda}F_{\kappa\rho}{F_\lambda}^\rho - 4R_{\nu\sigma}F_{\mu\rho}F^{\sigma\rho} -
2R^{\alpha\beta}F_{\alpha\mu}F_{\beta\nu} -
g_{\mu\nu}\nabla_\alpha\nabla_\beta({F^\alpha}_\rho F^{\beta\rho})
\right. + \\ \nonumber && \left. \qquad +\,
2\nabla_\alpha\nabla_\nu(F_{\mu\beta}F^{\alpha\beta}) -
\Box(F_{\mu\rho}{F_\nu}^\rho) \right) + \\ \nonumber
&+& c_6 \left(g_{\mu\nu}R^{\kappa\lambda\rho\sigma}F_{\kappa\lambda}F_{\rho\sigma}
- 6 F_{\alpha\nu}F^{\beta\gamma}{R^\alpha}_{\mu\beta\gamma} -
4\nabla_\beta\nabla_\alpha({F^\alpha}_\mu{F^\beta}_\nu) \right) +
\\ \nonumber
&+& c_7 \left(g_{\mu\nu}(F^2)^2 - 8F^2{F_{\mu}}^\sigma F_{\nu\sigma}\right) + \\ \nonumber
&+& c_8 \left(g_{\mu\nu}(\nabla_\kappa F_{\rho\sigma})(\nabla^\kappa
F^{\rho\sigma}) - 2(\nabla_\mu F_{\alpha\beta})(\nabla_\nu
F^{\alpha\beta}) - 4(\nabla_\alpha F_{\beta\mu})(\nabla^\alpha
{F^\beta}_\nu) \right.  + \\ \nonumber && \left. \qquad +\,
4\nabla_\alpha(F_{\nu\beta}\nabla^\alpha{F_\mu}^\beta) +
4\nabla_\alpha(F_{\nu\beta}\nabla_\mu F^{\alpha\beta}) -
4\nabla_\alpha({F^\alpha}_\beta\nabla_\nu {F_\mu}^\beta) \right) +
\\ \nonumber
&+& c_9 \left(g_{\mu\nu}(\nabla_\kappa F_{\rho\sigma})(\nabla^\rho F^{\kappa\sigma}) - 4(\nabla_\mu
F^{\alpha\beta})(\nabla_\alpha F_{\nu\beta}) - 2(\nabla_\alpha
F_{\beta\mu})(\nabla^\beta {F^\alpha}_\nu) \right.  + \\ \nonumber
&& \left. \qquad +\,
2\nabla_\alpha(F_{\nu\beta}\nabla^\alpha{F_\mu}^\beta) +
2\nabla_\alpha(F_{\nu\beta}\nabla_\mu F^{\alpha\beta}) -
2\nabla_\alpha({F^\alpha}_\beta\nabla_\nu {F_\mu}^\beta) \right)
\nxx
where we denoted $F^2 \equiv F_{\rho\sigma}F^{\rho\sigma}$.

\section{Corrections to the Reissner-Nordstr\"{o}m black hole in $D$ dimensions\label{app-ddim}}

The solution presented in Section \ref{sec-RN} can be easily
generalized to Reissner-Nordst\"{o}m black holes in $D$ spacetime
dimensions. (The unperturbed solution is presented in Refs.
\cite{Tangherlini} and \cite{Myers-Perry}.) The most general
spherically symmetric static metric in $D$ spacetime dimensions
has the form \n=={spherone} ds^2 = - e^{\nu(r)} dt^2 +
e^{\lambda(r)} dr^2 + r^2 d\Omega_{(D-2)}^2 \nxx where
\n=={sphertwo} \nonumber d\Omega^2_{(1)} = d\theta_0^2, \qquad
d\Omega^2_{(i+1)} = d\theta_i^2 + \sin^2\theta_i \,
d\Omega_{(i)}^2 \qquad 0 \leq \theta_0 \leq 2\pi, \quad 0 \leq
\theta_i \leq \pi \nxx so the metric for the coordinates
$(t,\,r,\,\theta_{D-3},\,\theta_{D-2},\,\ldots,\,\theta_0)$ is
\n=={spherthree} g_{\mu\nu} = \t{diag}\left(-e^{\nu(r)},\,
e^{\lambda(r)},\, r^2,\, r^2 \sin^2\theta_{D-3},\, \ldots,\,
r^2\sin^2\theta_{D-3}\cdots\sin^2\theta_2\sin^2\theta_1\right)
\nxx \n=={spherfour} \sqrt{-g} = r^{D-2} \, e^{(\nu + \lambda)/2}
\prod_{i=1}^{D-3}(\sin\theta_i)^i \nxx The corresponding
Christoffel symbols are \n=={spherfive} \Gamma^0_{00} =
\Gamma^0_{0k} = \Gamma^0_{11} =  \Gamma^0_{1k} = \Gamma^0_{kk'} =
0 \qquad\quad \Gamma^0_{01} = \frac{\nu'}{2} \nxx \n=={sphersix}
\nonumber \Gamma^1_{00} = \half \nu' e^{\nu-\lambda} \qquad
\Gamma^1_{01} = \Gamma^1_{0k} = \Gamma^1_{1k} = \Gamma^1_{kk'|k'
\neq k} = 0 \qquad \Gamma^1_{11} = \frac{\lambda'}{2} \qquad
\Gamma^1_{kk} = -\frac{e^{-\lambda}}{r} g_{kk} \nxx
\n=={spherseven} \nonumber \Gamma^k_{00} = \Gamma^k_{01} =
\Gamma^k_{0k'} = \Gamma^k_{11} = \Gamma^k_{1k'|k' \neq k} = 0
\qquad\quad \Gamma^k_{1k} = \frac{1}{r} \qquad\quad
\Gamma^k_\t{else}\;\; \t{not}\;\t{shown} \nxx where $k, k' =
2,\ldots,D-1$. The non-zero components of the Ricci tensor are
\n=={sphereight}
R^0_0 &=& -\frac{e^{-\lambda}}{2} \left(\nu'' + \frac{{\nu'}^2}{2} -\frac{\nu'\lambda'}{2} + (D-2)\frac{\nu'}{r}\right) \nonumber \\
R^1_1 &=& -\frac{e^{-\lambda}}{2} \left(\nu'' + \frac{{\nu'}^2}{2} -\frac{\nu'\lambda'}{2} - (D-2)\frac{\lambda'}{r}\right) \\
R^k_k &=& -e^{-\lambda} \left( \frac{(D-3)(1-e^\lambda)}{r^2} +
\frac{\nu'-\lambda'}{2r}\right) \nonumber \nxx \n=={sphernine} R
\,\, &=& -e^{-\lambda}\left(\nu'' + \frac{{\nu'}^2}{2} -
\frac{\nu'\lambda'}{2} + (D-2)(D-3)\frac{1-e^\lambda}{r^2} +
(D-2)\frac{\nu'-\lambda'}{r}\right) \nxx We can then write
\n=={spherten} \nonumber \frac{R^0_0 - R^1_1}{D-2} - R^k_k =
\frac{D-3}{r^2}(e^{-\lambda}-1) - \frac{\lambda' e^{-\lambda}}{r} =
\frac{\left(r^{D-3}(e^{-\lambda}-1)\right)'}{r^{D-2}} \nxx
\n=={sphereleven} \nonumber r^{D-3}(e^{-\lambda}-1) = \int dr\,
r^{D-2} \left( \frac{R^0_0 - R^1_1}{D-2} - R^k_k \right) \nxx
Assuming that the asymptotic behavior at $r \-> \inf$ is the
Schwarzschild solution, this becomes \n=={lambda-R-ddim}
e^{-\lambda} = 1 - \frac{2\kappa^2 M}{(D-2)\Omega_{D-2}\,r^{D-3}} -
\frac{1}{r^{D-3}}\int_r^\inf dr\, r^{D-2} \left( \frac{R^0_0 -
R^1_1}{D-2} - R^k_k \right) \nxx where \== \Omega_{D-2} =
\frac{2\,\pi^{(D-1)/2}}{\Gamma[(D-1)/2]} \xx is the area of the unit
sphere. Similarly, \n=={sphertwelve} \nonumber R^0_0 - R^1_1 =
-(D-2)\frac{e^{-\lambda}}{2}\frac{\nu' + \lambda'}{r} \nxx
\n=={nu-R-ddim} \nu = -\lambda + \frac{2}{D-2}\int_r^\inf dr\, r
\left( R^0_0 - R^1_1 \right) e^{\lambda} \nxx Einstein's equation
obtained from the action \n=={spherthirteen} S = \int d^Dx
\sqrt{-g}\,\frac{R}{2\kappa^2} + S_\t{matter} \nxx can be written as
\n=={Einstein-ddim} R_{\mu\nu} = \kappa^2 \left(T_{\mu\nu} -
\frac{T}{D-2}\,g_{\mu\nu} \right) \qquad\qquad T_{\mu\nu} =
-\frac{2}{\sqrt{-g}} \frac{\delta S_\t{matter}}{\delta g^{\mu\nu}}
\nxx and in our case $T = T^0_0 + T^1_1 + (D-2)T^k_k$. Then
\eqref{lambda-R-ddim} and \eqref{nu-R-ddim} become \n=={lambda-ddim}
e^{-\lambda} = 1 - \frac{2\kappa^2 M}{(D-2)\Omega_{D-2}\,r^{D-3}} -
\frac{2 \kappa^2}{(D - 2)\,r^{D-3}}\int_r^\inf dr\, r^{D-2}\,T^0_0
\nxx \n=={nu-ddim} \nu = -\lambda + \frac{2 \kappa^2}{D-2}
\int_r^\inf dr\,r (T^0_0 - T^1_1)\,e^\lambda \nxx The unperturbed
electrically charged solution is \n=={RN-metric-ddim}
e^\nu &=& e^{-\lambda} = 1 - \frac{2}{(D-2)\,\Omega_{D-2}}\, \frac{\kappa^2 M}{r^{D-3}} + \frac{1}{(D-2)(D-3)\Omega_{D-2}^2}\, \frac{\kappa^2 Q^2}{r^{2(D-3)}} \nonumber \\
E &=& \frac{Q}{\Omega_{D-2}\,r^{D-2}} \nxx We now consider an action
of the form \n=={RN-action-corr-ddim} S = \int d^Dx && \sqrt{-g}
\left( \frac{R}{2\kappa^2} - \frac{1}{4}F_{\mu\nu}F^{\mu\nu}
+ c_1\,R^2 + c_2\, R_{\mu\nu}R^{\mu\nu} + c_3\,R_{\mu\nu\rho\sigma}R^{\mu\nu\rho\sigma} + \right. \nonumber \\
&& \left. + \, c_4\,RF_{\mu\nu}F^{\mu\nu} +
c_5\,R^{\mu\nu}F_{\mu\rho}{F_\nu}^\rho +
c_6\,R^{\mu\nu\rho\sigma}F_{\mu\nu}F_{\rho\sigma} +
c_7\,(F_{\mu\nu}F^{\mu\nu})^2 \right. \nonumber \\
&& \left. + \, c_8\,(\nabla_{\mu} F_{\rho\sigma})(\nabla^{\mu}
F^{\rho\sigma}) + c_9\,(\nabla_{\mu}
F_{\rho\sigma})(\nabla^{\rho}F^{\mu\sigma}) \right) \nxx By the same
procedure as described in the main text, we consider the corrections
to the effective $T_{\mu\nu}$ based on the equations in Appendix
\ref{app-Tmn}, and obtain the mass-charge relation for extremal
black holes \n=={mass-ddim} \frac{D-3}{D-2} \frac{\kappa^2 M^2}{Q^2}
&=& 1 - \frac{2(D-3)}{(D-2)(3D-7)}
\left(\frac{(D-2)(D-3)\Omega_{D-2}^2}{\kappa^2 Q^2}\right)^{1/(D-3)}
\times \\ \nonumber &\times& \left[ (D-3)(D-4)^2\,\kappa^2 c_1
+(D-3)(2D^2-11D+16)\,\kappa^2 c_2 + \right. \\ \nonumber && \left.
+\, 2(2D^3-16D^2+45D-44)\,\kappa^2 c_3 + 2(D-2)(D-3)(D-4) c_4  +
\right.
\\ \nonumber && \left. +\, 2(D-2)(D-3)^2 c_5 + 2(D-2)(D-3)^2 c_6 +
4(D-2)^2(D-3)\frac{c_7}{\kappa^2} - \right. \\ \nonumber && \left.
-\, 2(D-2)(D-3)^2 c_8 - (D-2)(D-3)^2 c_9 \right] \nxx It is
convenient to choose the normalization $\kappa^2 = (D-2)/(D-3)$, and
then \n=={mass-ddim-norm} \frac{M^2}{Q^2} = 1 &-& \frac{2}{3D-7}
\left(\frac{(D-3)\,\Omega_{D-2}}{Q}\right)^{2/(D-3)} \times \\
\nonumber &\times& \left[
(D-3)(D-4)^2 c_1
+ (D-3)(2D^2-11D+16) c_2 + \right. \\ \nonumber && \left.
+ 2(2D^3-16D^2+45D-44) c_3
+ 2(D-3)^2(D-4) c_4
+ 2(D-3)^3 c_5 + \right. \\ \nonumber && \left.
+ 2(D-3)^3 c_6
+ 4(D-3)^3 c_7
- 2(D-3)^3 c_8
- (D-3)^3 c_9
\right]
\nxx


\end{document}